\documentclass{ecai} 

\usepackage{latexsym}
\usepackage{amssymb}
\usepackage{amsmath}
\usepackage{amsthm}
\usepackage{booktabs}
\usepackage{enumitem}
\usepackage{graphicx}
\usepackage{color}
\usepackage{balance}

\usepackage{multirow}
\usepackage{threeparttable}
\usepackage{array} 
\usepackage{algorithm}
\usepackage{algorithmic}
\usepackage{makecell}
\usepackage{balance}

\usepackage[font=small]{caption}

\newcommand{\BibTeX}{B\kern-.05em{\sc i\kern-.025em b}\kern-.08em\TeX}

\begin{document}


\begin{frontmatter}


\paperid{6} 


\title{Value-Enriched Population Synthesis: Integrating a Motivational Layer}

\author[A]{\fnms{Alba}~\snm{Aguilera}\orcid{0009-0003-5336-8570}\thanks{Corresponding Author. Email: aaguilera@iiia.csic.es}}
\author[B]{\fnms{Miquel}~\snm{Albertí}\orcid{0009-0005-1666-8421}}
\author[A]{\fnms{Nardine}~\snm{Osman}\orcid{0000-0002-2766-3475}} 
\author[C]{\fnms{Georgina}~\snm{Curto}\orcid{0000-0002-1320-3873}} 

\address[A]{Artificial Intelligence Research Institute (IIIA-CSIC), Barcelona}
\address[B]{Universitat de Barcelona, Barcelona}
\address[C]{University of Notre Dame, Notre Dame, USA}

\begin{abstract}
    In recent years, computational improvements have allowed for more nuanced, data-driven and geographically explicit agent-based simulations. So far, simulations have struggled to adequately represent the attributes that motivate the actions of the agents. In fact, existing population synthesis frameworks generate agent profiles limited to socio-demographic attributes. In this paper, we introduce a novel value-enriched population synthesis framework that integrates a motivational layer with the traditional individual and household socio-demographic layers. Our research highlights the significance of extending the profile of agents in synthetic populations by incorporating data on values, ideologies, opinions and vital priorities, which motivate the agents' behaviour. This motivational layer can help us develop a more nuanced decision-making mechanism for the agents in social simulation settings. Our methodology integrates microdata and macrodata within different Bayesian network structures. This contribution allows to generate synthetic populations with integrated value systems that preserve the inherent socio-demographic distributions of the real population in any specific region.

\end{abstract}

\end{frontmatter}

\section{Introduction}

Agent-based simulations are now widely used in interdisciplinary research, especially to support policy-making in social contexts. When applied to real-life domains, these simulations are evolving towards more nuanced, data-driven models that accurately reflect the complexities of socio-environmental systems. During the outbreak of the COVID-19 pandemic, the importance of accurate simulations in policy-making scenarios became starkly evident. An important body of studies emerged focusing on simulating disease outbreaks and public health interventions, such as lockdowns or mask mandates. These types of models (along with many others focusing on issues like gentrification, spatial inequality or urban ageing) require a reliable representation of agents, the interactions between them and their environment~\cite{yossef2023social}.

In such models, the interactions and behaviour of agents are strongly determined by their profile, which usually comprises specific demographic attributes. These attributes must closely resemble the characteristics of the human population they simulate. Ideally, this information would be sourced from census data; however, due to data privacy constraints, only certain open-source data can be employed. To achieve a simplified yet representative depiction of the population in a specific region, population synthesis methods are used. In this context, the most important challenge is to close the gap between the generated population and the actual one~\cite{chapuis2019brief}. Traditional synthetic populations generate profiles limited to demographic and socioeconomic information~\cite{fabrice2021comparing}. We argue that there is an additional type of data, referred to as motivational attributes in this article, that can enrich population synthesis frameworks, contributing to reflect the complexities of social interactions and decisions. This data includes cognitive and cultural attributes such as values, ideologies, opinions and vital priorities that have a direct impact on the agents' behaviours~\cite{Schwartz2012}. We argue that the behaviour of agents, which can either be learnt via machine learning techniques or modelled through mathematical decision-making architectures, can be enhanced by relying on these motivational attributes. Our approach opens the door to the incorporation of numerous frameworks that define and quantify both individual and collective values, such as the cross-cultural Schwartz Theory of Basic Values~\cite{Schwartz2012}, the evolution of values in line with economic security (or post-materialistic), described by Inglehart~\cite{inglehart2005modernization}, the capability approach to human development~\cite{Sen2001}, multiculturalism and the struggle for recognition~\cite{Taylor2009, Honneth1996}, as well as the impact of prejudices~\cite{Allport1954,Xu2014}. In this article, we have focused on incorporating the data obtained by surveys following the frameworks defined by Schwartz's and Inglehart's~\cite{wvs, evs,catvs, bvs}.  By incorporating this data in a population synthesis framework, we are contributing to a line of research that increases the realism of social simulations and motivating researchers in the social domains to use empirical data in their population generation processes~\cite{wan2019sync, chapuis2022generation}. 
Motivational attributes are not randomly nor uniformly distributed across different regions or social groups, but are deeply correlated with the socioeconomic development of a region~\cite{inglehart2005modernization}. ~\cite{cohen2001cultural} We propose to address this complexity by extending the current demographic-based profile of agents in synthetic populations with the addition of a motivational layer that automatically adapts to the specific region of the case study. The approach therefore facilitates the replication of the model in different regions of the world, working towards policy-oriented research that includes both the Global North and the Global South~\cite{de2024recommendations}. This motivation layer contains information about social aspects of the agents' profiles, reflected in the surveys in scope, that potentially motivate behaviour. Decision-making techniques can then rely on these parameters to fine-tune data-driven outcomes, contributing to the "approach to reality" decision-making research direction~\cite{LIANG2022108982}. However, sources of motivational data are often limited in size and do not cover the whole population's socio-demographic attributes. They require meticulous integration with other data sources to ensure that the synthetic population, which includes both socio-demographic and motivational characteristics, is sufficiently representative within the desired geographical scope. 

\begin{figure}[ht]
    \centering
    \includegraphics[width = \columnwidth]{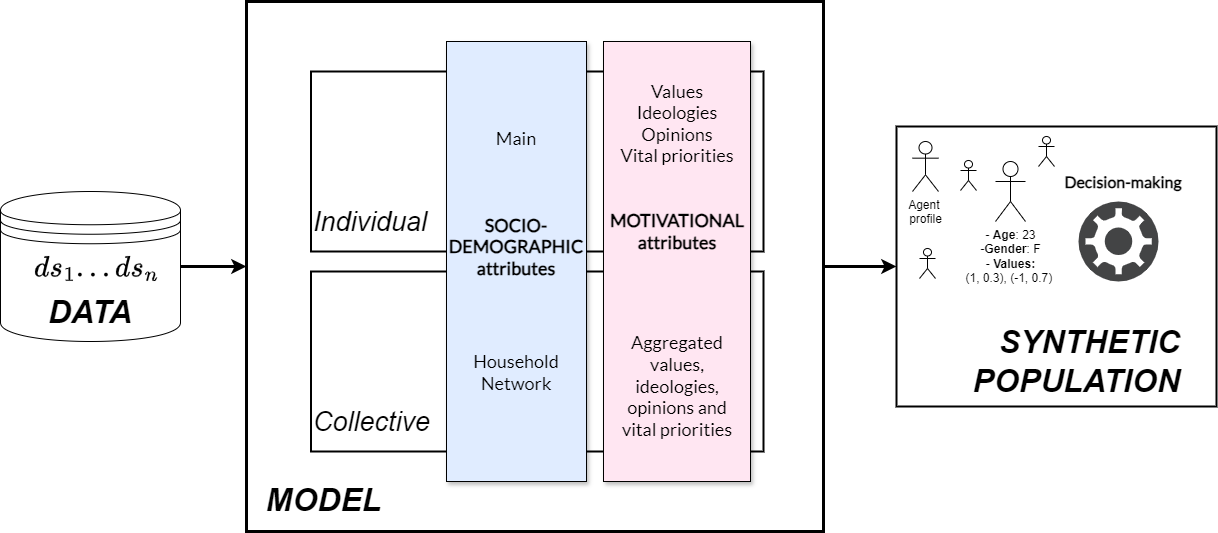}
    \caption{Workflow scheme. From diverse data sources we extract personal profile attributes and classify them into socio-demographic and motivational layers, represented in blue and pink, at both the individual and collective levels. This information is used to generate comprehensive agent profiles in synthetic populations that help decision-making architectures tune behaviour.} 
    \label{fig: conceptual}
\end{figure}

To bridge this gap, we propose a novel value-enriched population synthesis framework that integrates diverse data sources to generate a population that goes beyond the traditional individual and household layers. Through value learning from existing data, we explore the dependencies between attributes that accurately represent the underlying motivational ones in the population. In particular, by following our methodology, we can generate a synthetic population with a (1) socio-demographic layer and a (2) motivational layer at both the individual and collective levels (see Fig.\ref{fig: conceptual}). Our approach aims to be replicable and scalable to a series of case studies. Therefore, we use Bayesian networks that can either be expanded or reduced depending on the data requirements of the study. To overcome data scarcity issues, which is one of the main challenges of population synthesis today, our approach is able to process both macrodata and microdata. We provide a proof of concept for the framework with a use case in the city of Barcelona. We present how we can potentially generate a synthetic population that uniquely characterizes individuals (i.e. their demographic profile along with their value system, ideologies, opinions, worries, priorities, etc.) in a representative manner. 

The paper is organized as follows: Section~\ref{sec: relworks} underlines the novelty of our work by exploring the state of the art in value-enriched population synthesis. In Section~\ref{sec: background}, we describe the proposed data model and the integration of values into the framework, while in Section~\ref{sec: method}, we explain the formulation of our population synthesis proposal. Section~\ref{sec: usecase} presents a comprehensive application of the framework through a use case for the city of Barcelona. Finally, we conclude with insights into the main implications, limitations, and avenues for future research in Section~\ref{sec: concl}.

\section{Related Work}
\label{sec: relworks}

Synthetic populations are essential to develop useful applications in an ethical way that does not compromise individual privacy. Numerous works have generated open-source synthetic populations; for the UK~\cite{UK2022}, the US~\cite{US2009}, Canada~\cite{CANADA2023}, Ile-de-France (France)~\cite{FRANCE2021}, Tallinn (Estonia)~\cite{Tallin} or some Australian cities~\cite{Australia} (e.g. Sydney, Melbourne and Brisbane), to name a few. These studies rely on two essential components: (i) data sources, such as publicly available microsamples, surveys or government databases, and (ii) population synthesis techniques, which are being constantly refined to match the complex interrelationships between the agent attributes. However, none of these studies have yet incorporated an agents' motivational layer into their frameworks. Adding such a layer could be highly beneficial for models aiming to more accurately represent the complexities of the physical and social environments as a single fabric, a concept recently referred to as Social Urban Digital Twins (SUDTs)~\cite{yossef2023social}. Let us review the main established data sources and synthesis techniques.

Handling data is perhaps the hardest challenge for population synthesis nowadays, especially due to the variability in format, size, level of detail and disaggregation of data sources across different regions. There are a lot of initiatives that work towards data availability and harmonization by offering detailed socio-demographic information, such as IPUMS (International Public Use Micro Samples)~\cite{IPUMS}. In our case, we are particularly interested in existing surveys on values. These are international research programs devoted to the scientific study of social, political, economic, religious and cultural values of people in the world. Similar surveys exist at the national level, offering practically the same sort of information along different territorial units. For instance, The World Value Survey (WVS)~\cite{wvs}, the Europe Value Study (EVS)~\cite{evs}, the European Social Survey (ESS)~\cite{ess}, the Catalonian value survey~\cite{catvs} and the Barcelona value survey~\cite{bvs} contain almost completely harmonized data along continent, country, region, municipality and district. These data sources provide a link between socio-demographic and motivational attributes at different geographic scopes. 

The literature categorizes the techniques used to generate synthetic populations into three main groups: synthetic reconstruction, combinatorial optimization, and statistical learning. The choice among the different techniques largely depends on the characteristics of the available data~\cite{fabrice2021comparing}. The first category uses deterministic algorithms that fit and allocate fractions of individuals and/or households to a region, while the second one attempts to reach an optimized solution by randomly drawing from the microsample while minimizing differences in marginals. More recently, researchers have adopted a probabilistic framework instead of a deterministic one, which corresponds to the third category and is the primary focus of this study. This approach searches for the joint distribution of all attributes using partial views available in the data. Within this category, we settle for Bayesian networks because of the main advantages this approach offers: it facilitates replicability (clear and graphical interface) and scalability (easy parallelization for large-size samples). Although many other methods have technically surpassed the capability of sampling a synthetic population, we prioritize successfully merging different data sources rather than the technical accuracy of the method. The foundational application of Bayesian network synthesis can be traced back to~\cite{sun2015bayesian}, followed by other contributions~\cite{ZHOU2022101717, sallard2023travel, ilahi2019integrating}, which constitute our technical starting point. 

This study aims to contribute to population synthesis for agent-based models under a social lens. The main novelty of our work is the aggregation of a motivational layer, sourced from social surveys, into the synthetic population. The use case developed for the city of Barcelona is intended to advance the enhancement of the Aporophobia Agent-Based Model (AABM) project~\cite{AMPM, Aguilera2024}. The successful addition of motivational attributes into the population signifies the possibility of adjusting the needs-based model~\cite{Dignum2020} decision-making architecture with real-world data.

\section{Proposed Data Model}
\label{sec: background}
The motivation behind our model is to use diverse data sources to generate comprehensive agent profiles that help decision-making architectures simulate agents' behaviour. In particular, we aim to extend the individual and collective socio-demographic attributes of agents with motivational ones, providing a deeper understanding of the individual’s value system that contributes to the agents' actions~\cite{Schwartz2012}. The workflow is illustrated in Fig.~\ref{fig: conceptual}. From the available data sources, denoted as $ds_1, ..., ds_n$, we extract personal attributes and categorize them into two main layers: (i) socio-demographic and (ii) motivational, at both the individual and collective levels. Due to the high dimensionality, particularly of motivational attributes, we further divide the attributes within each layer into types. This allows having an organized overview of the information integrated into the synthetic population.

Socio-demographic attributes, represented in blue, cover the available social, demographic and economic characteristics that define an individual inside a population, a household and a social network. This layer includes "main" attributes at the individual level such as age, gender and nationality, as well as "household" and "network" attributes at the collective level, such as the number of people one lives with, the number of children, detailed information about the household assets and the number of friends. On the other hand, the motivational layer, represented in pink, encompasses attributes that directly impact or motivate behaviour. For this article, they are regrouped into four types: values, ideologies, opinions and vital priorities. This layer includes individual and collective information about prevalent values measured following Inglehart's or Schwartz's theory. Additionally, it includes other motivational attributes such as the alignment with various ideologies, the perceptions on political or economic situations or institutions, the points of view on controversial topics and the 
vital priorities given to certain aspects of one's life (see Table~\ref{tab: variables} for further details). These individual motivational attributes can become collective (or consensus) attributes when aggregated or averaged along a region~\cite{lera2024aggregating}. 

\subsection{Values within the Motivational Layer}

 Information extracted from social surveys on values is used to construct the motivational layer of the synthetic population. The surveys are structured along thematic sub-sections covering diverse topics such as vital life priorities, societal well-being, social values and attitudes, religious values, political behaviour and ideology, cultural and national identity, and opinions towards minorities, climate change, migration, etc~\cite{ESS2023}. 

 The sections directly related to values in the surveys are linked to the two major human values theories: Inglehart's and Schwartz's. However, keep in mind that surveys can be linked with other theoretical frameworks (such as the ones mentioned in the introduction). Some motivational data sources, such as the ESS~\cite{ess}, use Schwartz's portrait values questionnaire (PVQ-21)~\cite{PVQ21} to measure the ten fundamental values. In contrast, sources such as WVS~\cite{wvs} or EVS~\cite{evs}, use other methods like Inglehart's materialism/post-materialism (MPM) index~\cite{MPM, MPMmap}. Other regional sources (such as Catalonia's and Barcelona's value surveys) use a combination of both approaches. In all the approaches, respondents' choices are used to quantify their value preferences. 

\begin{figure}[h]
    \centering
    \includegraphics[width=\columnwidth]{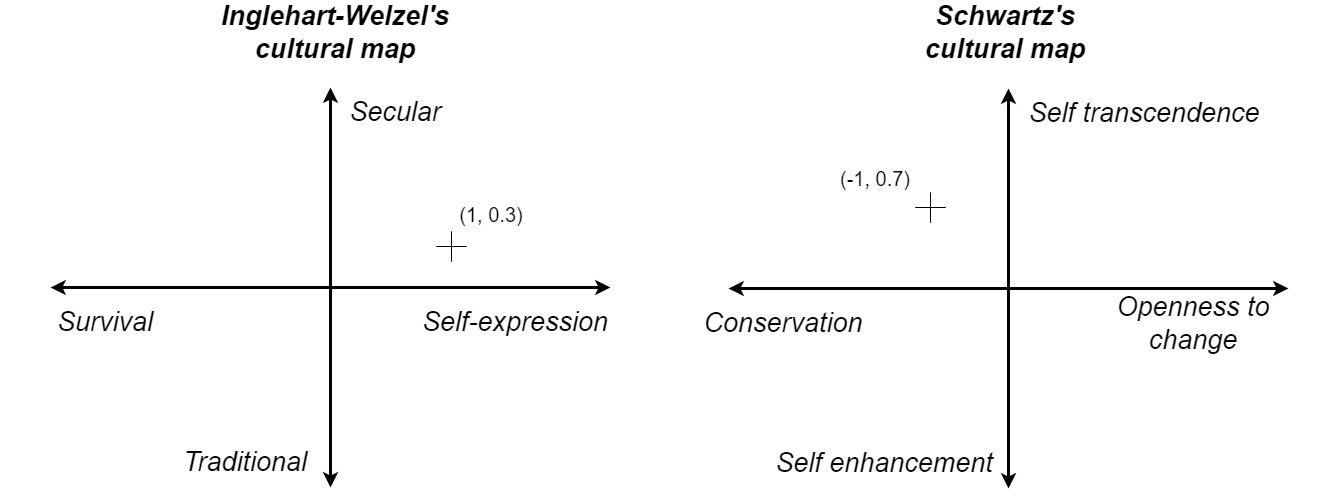}
    \caption{Simplified depiction of the Inglehart-Welzel's and Schwartz's cultural maps. The points represent a symbolic value preference measured with each value theory along different dimensions.} 
    \label{fig: culturalmap}
\end{figure}

The resulting value preference can be depicted with cultural maps across two predominant dimensions that encompass different values. The Schwartz map~\cite{Schwartzmap} includes the dimensions "conservation versus openness to change" and "self-enhancement versus self-transcendence", while the Inglehart-Wezel map~\cite{Inglehartmap} features the dimensions "traditional versus secular-rational values" and "survival versus self-expression values." Both maps are represented schematically in Fig.~\ref{fig: culturalmap}, where each specific point indicates a personal value preference. 
\vspace{-0.3cm}
\section{Formulation}
\label{sec: method}
In the context of population synthesis under a probabilistic lens, the aim is to infer the underlying joint probability distributions of data, denoted as P$(X_i, .., X_n)$. The random variables $X_i, .., X_n$ are the agent's profile attributes, where each variable $X_i$ is capable of assuming various states $x_k$. Bayesian networks can embed this joint distribution through two fundamental components: structure and parameters. The structure $S$ captures the dependencies among these variables by connecting them. The parameters $\delta$ determine the conditional probability agents have of being assigned a certain attribute given that they have been assigned another one. 

Given a set of observational data $\mathcal{D}$, one can either learn the parameters $\delta$ when the structure $S$ is known or learn both the structure $S$ and parameters $\delta$. These processes are known as parameter learning and structural learning, respectively. Our approach employs both methods by (1) crafting an intuitive network based on prior knowledge (knowledge-based model) and (2) using heuristic search techniques to learn an optimal structure (learnt model). These models are then used to sample from the joint probability distribution, obtaining a synthetic dataset where each column is a socio-demographic or motivational attribute and each row uniquely characterizes one individual. The general workflow of a population synthesis process comprises three main steps: (i) data preparation, (ii) model selection and (iii) model validation. 

\subsection{Data Preparation}
 Our model is adapted to handle diverse data. We typically encounter two primary data forms: (1) macrodata, which provides information on separate attribute sets but lacks detail on their interdependencies, and (2) microdata, from which we can identify comprehensive interdependencies among attributes. Macrodata can provide information about the conditional probabilities of one, two, or three attributes at most. Relying solely on these conditionals or marginals to produce a representative sample of agents is not feasible, so it is often used as marginal constraints to control the population synthesis process. On the other hand, microdata offers richer insights into the interdependencies between the available attributes (contains their joint distribution) and can be efficiently used to generate a sample. Nevertheless, this microdata can be outdated, lack the needed granularity or contain numerous missing values. 

Before data can be integrated into the model, it is essential to select, clean and harmonize the information in it. On one side, the selection of attributes largely depends on the quality of the data. If data on an attribute is substantially incomplete, which is a common situation regarding motivational data, the attribute may be discarded. However, for partially complete data, missing values can be handled either by direct elimination or through imputation techniques~\cite{osman2018survey}. Data harmonization is then required to make diverse datasets compatible and consistent by standardizing and normalizing their values and integrating them into a single, cohesive dataset.

\subsection{Model Selection}
\label{sec: model selection methodology}
Following data preparation, the next crucial step is constructing the model that integrates all the selected attributes. These attributes can be organized according to our proposed data model, illustrated in Fig.~\ref{fig: conceptual}, where attributes are classified within socio-demographic and motivational layers at both the individual and collective levels. Typically, we can associate each layer with one or several data sources that contain the corresponding household, network or motivational attributes. We denote the set of attributes within each data source $ds \in \mathcal{D}$ as $X^{ds}_{i}$, where $i = 1, ..., m_{ds}$ and $m_{ds}$ is the total number of attributes selected from a specific data source. 

We aim to connect the profile attributes in $\mathcal{D}$ with a structure $S$ and associated parameters $\delta$, representing conditional probabilities between them. In the context of Bayesian networks, various strategies exist for structuring and parameterization, which can differ when dealing with microdata or microdata. Structural learning and parameter learning are straightforward processes for microdata using the common Bayesian network libraries. However, the joint distribution is not attainable for macrodata; only the conditional probabilities of the attributes present separately in each data chunk can be obtained by applying Bayesian estimators of one's choice~\cite{pgmpylearningstructure}. 

To address this limitation, we create two models: one that works solely with microdata (learnt model) and another that can integrate macrodata (knowledge-based model). The main difference between them is how the structure is obtained. While the learnt model incorporates dependencies between attributes directly from data, the knowledge-based model has a predefined structure based on prior knowledge. This combination of models allows for the enrichment of less comprehensive datasets with the robustness of more complete or up-to-date ones. The precise application of the models is detailed in the application through a use case (Section~\ref{sec: usecase}).

In either of the cases (learnt or crafted), for the attributes in each data source $ds \in \mathcal{D}$, we define a structure $S_{ds}$ and parameters $\delta_{ds}$ that connect the attributes within that data source. In other words, $S_{ds}$ is a graph with vertices $\{X^{ds}_i\}_{i=1,...,m_{ds}}$ and directed edges $\{(X^{ds}_i, X^{ds}_j)\}_{i,j\in\{1,...m_{ds}\}}$, while $\delta_{ds}$ is a set of conditional probability tables $\left\{ P(X^{ds}_i \mid X^{ds}_j)\right\} $ that characterize the dependencies between the attributes described by the $S_{ds}$ structure. 

\begin{figure}[htb]
    \begin{center}
    \includegraphics[width= \columnwidth]{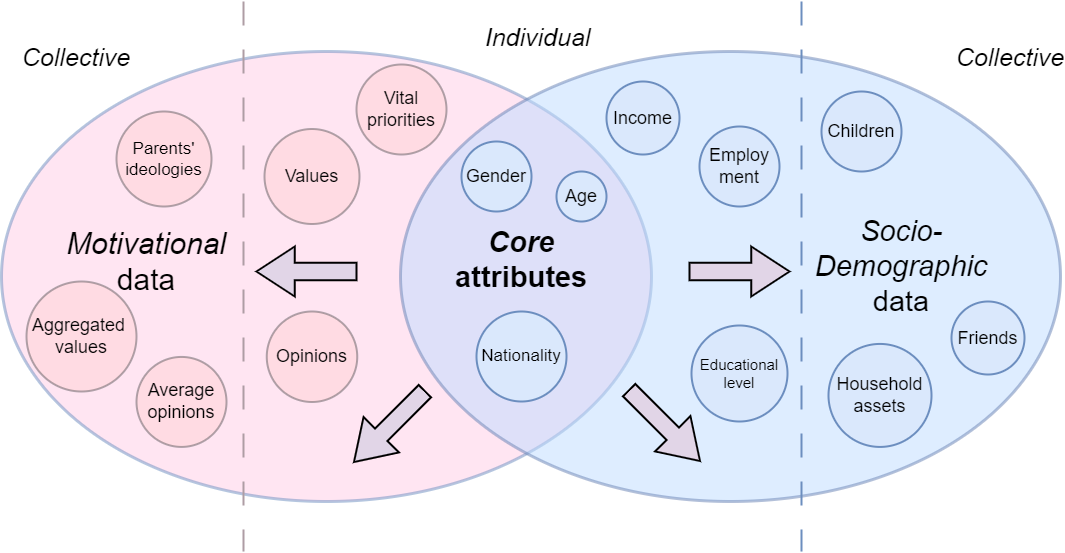}
    \end{center}
    \caption{Representation of the intersection between a set of observational data that contains socio-demographic and motivational information at both the individual and collective levels. As signalled by the edges, from the set of \textit{core} attributes, our procedure allows us to derive the structures or dependencies of all the remaining personal attributes.}
    \label{fig: modelscheme}
\end{figure}

The functioning of our model relies on the combination of various datasets that share a common intersection. We will refer to this set of intersecting variables as the \textit{core} attributes. As all data sources (including surveys, governmental sources and public use microsamples) typically have the same common information integrated, the \textit{core} attributes always correspond to either just individual or individual and collective socio-demographic attributes (e.g. gender, age, nationality, etc.) as shown in Fig.~\ref{fig: modelscheme}. As signalled, by the edges, the rest of the attributes in the socio-demographic and motivational layers are obtained from the \textit{core} attributes.

Let $S_{core}$ refer to the structure connecting the core attributes, learnt or craft from the richest dataset available $ds_{rich}$. The richest dataset is selected based on various criteria, such as being the largest, the most disaggregated, or the most up-to-date, depending on the specific preferences and requirements of one's study for accurately representing the socio-demographic characteristics of the population. We can use the \textit{core} attributes as a binding factor for the remaining structure, which includes all $X^{ds}_i \notin core$. We do so by fixing $S_{core}$ when learning the other structures, avoiding the overwriting of previously learnt $\delta_{core}$ parameters. This approach ensures that the motivational and remaining socio-demographic attributes are linked with the \textit{core} attributes while respecting the probability distributions in each data source. The general procedure to learn the structure and parameters connecting all the attributes is the following 
\begin{algorithm}
\caption{Procedure to merge attributes contained in different data sources.}
\label{alg:merge_profiles}
\begin{algorithmic}[1]
    \STATE Find the \textit{core} attributes present in all data sources $ds$. 
    \STATE Detect the richest dataset available $ds_{rich}$.
    \STATE Craft $S_{core}$ or learn it from $ds_{rich}$. 
    \STATE Learn or computing $\delta_{core}$ from $ds_{rich}$. 
    \STATE Fix $S_{core}$ while learning or crafting $S_{ds}$ for all $ds$ with  $X^{ds}_i \notin core$. 
    \STATE In the case of learning, eliminate any introduced edge that conflicts with $\delta_{core}$, i.e. edges between \textit{core} attributes or edges that go from other data sources to \textit{core} attributes. 
    \STATE Learn or compute $\delta_{ds}$  from each $ds \neq ds_{rich}$ for the newly incorporated variables $X^{ds}_i$ $\notin core$.
\end{algorithmic}
\end{algorithm}

The final structure is a composition of the structures defined for each data source. The set of nodes is connected through a set of edges, which connect attributes within the same data sources, except from the ones going from the core to other data sources.

\subsection{Model Validation}

Model validation is the process of ensuring the accuracy of the model by comparing the generated synthetic population with the original data. Researchers choose the validation metrics based on the constraints they want to fulfil. In general, these constraints try to ensure that the probability distributions that will eventually be used in population synthesis (whether marginal, conditional, joint, or partially joint) are close enough to the distributions of the real data. 

As we are considering an ensemble of different types of data, i.e. macrodata, microdata or a combination of both, we propose to adapt the validation to these different types. Various metrics can be used, such as: (i) Wasserstein distances~\cite{wasserstein} for evaluating marginal distributions (to be used with macrodata), and (ii) regression lines or SRMSE, described in~\cite{sun2015bayesian}, for evaluating joint distributions (to be used with microdata). The first metric intuitively shows how different two probability distributions are. Concretely, it measures the minimum amount of "work" or effort required to transform one distribution into another. The lower the distance, the closer the distributions. The second comparative approach is regression lines, which help visualize the fit of the synthetic population to the weighted microsample through a frequency plot, where the frequencies of every unique variable combination in the two datasets are plotted against each other. A perfect match is represented by a line of best fit with zero intercept, unit slope, and a correlation coefficient value of one.

\section{Application through Use Case}
\label{sec: usecase}
 
Following the steps defined in Section~\ref{sec: method}, one can apply our value-enriched population synthesis framework for a specific set of data sources in a region. Once the data is prepared, the models can be created and validated against the original data. The metropolitan area of Barcelona is selected as a proof of concept for our framework. We generate a synthetic population with agents aged from 15 to 74 years old, as motivational data (from existing surveys) is only available for people in that age range.

We use Python's Bayesian network library \textit{pgmpy}~\cite{pgmpy}, which allows for the direct implementation of parameter learning and structural learning techniques. All the project materials can be accessed from the corresponding GitHub public repository. 
\footnote{https://github.com/albaaguilera/Population-Synthesis} 

\subsection{Data Preparation}

After a thorough analysis of the available data for our selected region, working in close collaboration with local governmental organizations (such as the Open Data department and the Opinion Studies Center of the Government of Catalonia) and non-profit organizations (such as fundació Bofill, which focuses on promoting critical knowledge through education-related studies), we decided upon a set of data sources. The primary data sources considered (and their most updated year of coverage) are: OpenData (2022)~\cite{opendata}, IPUMS (2011)~\cite{IPUMS}, Panel fundació Bofill (2012)~\cite{Panel}, Barcelona values survey (2021)~\cite{bvs} and Catalonia values survey (2023)~\cite{catvs}, which are presented in further detail in Table~\ref{tab: data}. 
\begin{table}[h!]
    \centering
    \caption{Selected data sources after attribute selection and harmonization. The size of the data sources, represented by the number of individuals, is listed, along with the maximum geographic scope, the type of data and the models to which the data is fed.}
    \begin{tabular}{p{0.3cm}  >{\centering\arraybackslash}p{0.8cm} >{\centering\arraybackslash}p{1.4cm} >
    {\centering\arraybackslash}p{1.4cm}  > {\centering\arraybackslash}p{1cm} >{\centering\arraybackslash}p{1.2cm}} 
    
        \textbf{No.} & \textbf{Source} & \textbf{Size} & \textbf{Scope} & \textbf{Type} & \textbf{Model} \\ \hline
        $ds_1$ & OpenData & 1,600,000 & Neighborhood & Macrodata & Knowledge-based \\ \hline
        $ds_2$ & IPUMS & 120,000 & Municipality  & \multirow{4}{*}{\makecell{\\ \\ \\ \\ Microdata}} & \multirow{4}{*}{\makecell{\\ \\ \\ Knowledge-\\based and\\ learnt}} \\ \cline{1-4}
        $ds_3$ & Panel & 1,500 & Census section  & & \\ \cline{1-4}
        $ds_4$ & Bcn values survey & 1,300 &  District  & & \\ \cline{1-4} 
        $ds_5$ & Cat values survey & 3,100 &  Region  & & \\ \hline
    \end{tabular}
    \label{tab: data}
\end{table}

OpenData~\cite{opendata} is a governmental database, updated annually, that provides socio-demographic macrodata for the entire population of Barcelona up to the neighbourhood level. As a macrodata source, only the knowledge-based model supports it. IPUMS, Panel and value surveys' data are microdata sources, used to feed the learnt and knowledge-based models. IPUMS~\cite{IPUMS} contains detailed socio-demographic information from the Spanish census at both the individual and collective levels. Panel fundació Bofill's data~\cite{Panel} originates from a longitudinal survey aimed at exploring social inequalities in Catalonia, conducted across households up to the census section level. The Barcelona values survey~\cite{bvs} and the Catalonia values survey~\cite{catvs}, conducted every two years up to the district and municipality level, investigate ideological, ethical, or attitudinal questions to understand the prevailing value system of the population. We acknowledge that combining data from different years involves a significant assumption, as the populations described may have changed over time. However, the validation process is designed to address this assumption by evaluating the accuracy of the models against all the available data sources. This approach helps us ensure that, despite the temporal differences in the data, the synthetic population does not deviate from the actual characteristics of the population in the region.

\begin{table*}[th]
\caption{Summarized classification, definition and description of the selected profile attributes for population synthesis. The attributes are organized into types within the socio-demographic and motivational layers. Note that this simplified table does not specify the $70$ attributes comprised in the agent's profile. }
\label{tab: variables} 
\renewcommand{\arraystretch}{1.35}
    \begin{tabular}{p{2cm} p{1.65cm} p{1.5cm} p{5.3cm} p{5.4cm}}

\textbf{Layer} & \textbf{Type} & \textbf{Attribute} & \textbf{Definition} &  \textbf{Description}\\ \hline
\multirow{13}{*}{\makecell{Socio-\\demographic}} & \multirow{7}{*}{Main } & D$^{1}$ & District&  Territorial unit\\ 
                                   &                           & G & Gender &  Female or male\\ 
                                   &                           & A & Age &  $0 - 100$ years old by groups of $10$\\ 
                                   &                           & N & Nationality &  Spain / rest of EU / rest of the world\\ 
                                   &                           & E & Educational level &  Last educational level attainment \\ &
                                    & U & Unemployment &  Registered employed or unemployed\\ 
                                   &                            & I & Income &  Monthly amount of income\\ \cline{2-5}
                                     &  \multirow{2}{*}{Household} & Hr & Number of people you live with &  Number from 0 to 4 or more\\  &                           & Ch & Children in the household &   No children / one or more children\\  
 \cline{2-5}
                              & \multirow{2}{*}{Network} & Fr & Number of friends. & Number from 0 to 3 or more \\ &       
         & $X_{Fr}$ \footnote{} & Friends' main demographic attributes &  Friends' age, gender, educational level and nationality\\  \hline
\multirow{4}{*}{Motivational} & Values & \multirow{2}{*}{} &  Inglehart's materialist/post-materialist index &  Degree of materialism, mixed values or post-materialism ($1 - 7$) \\ & &
                              &  Alignment with Schwartz's fundamental values & Degree of agreement or disagreement with the $10$ fundamental values ($1 - 5$) 
                              \\     \cline{2-5}
    
                              & Ideologies & \multirow{3}{*}{} & Individual's and parents' ideology &  Political spectrum ($1-8$) \\ & &
                              & Alignment with capitalism, socialism, communism and political independence movements & Degree of agreement: agreement, disagreement and indifference ($1-3$)\\ & & & Alignment with feminism, ecologism, multiculturalism and religion &  Degree of agreement: agreement, disagreement and indifference ($1-3$)\\  \cline{2-5}
                              & Opinions & \multirow{3}{*}{} & Interest on politics, sports, culture, etc. &  Degree of interest ($1-4$) \\ & &
                              & View on controversial topics such as immigration, squatting, sustainability, etc. &  Multiple options varying with topic \\ & & & Confidence in the police, the state, the government, the church, the people, etc. &  Degree of confidence ($1-4$)\\  \cline{2-5} 
                              & Vital priorities  &  & Importance given to or satisfaction provided by certain aspects of one's life: family, friends, work, personal time and studies &  Degree of importance or satisfaction ($1-10$)\\  \hline

    \end{tabular}
\begin{tablenotes}
\footnotesize
\item[] $^1$ Territorial unit interchangeable for other geographic scopes present in the data sources (e.g. municipality "M", census section "CS" or neighbourhood "N"). 
\item[] $^2$ $X_{\text{Fr}}$ refers to the set of variables describing the demographic profile of the individual's friends (e.g. $G_{1}$ being the number one friend's gender). 
\end{tablenotes}
\end{table*}
Given the set of data sources, a thorough cleaning, harmonization and selection of attributes need to be performed. In our case, we resort to the direct elimination of missing values and the establishment of simplified and standardized states $x_k$ for each variable to ensure consistency. Additionally, the territorial unit node is designed to be flexible, allowing our application to adapt to various spatial scales.

\subsection{Model Selection}

A summarized classification, definition and description of the selected attributes' is provided in Table~\ref{tab: variables}. The models encompass over seventy variables, a number that can either be reduced or extended to narrow or widen the synthetic agent's profile. We could include as much information as the data sources allow (e.g. workplace sector, extracurricular activities, information about the use of time, the perception of one's health state or social class, detailed information on the household assets, etc). As explained in Section~\ref{sec: method}, we differentiate between two models: (1) the learnt model (Section~\ref{sec: S}) and (2) the knowledge-based model (Section~\ref{sec: H}).

\subsubsection{Learnt Model}
\label{sec: S}

\begin{figure*}[t]
    \centering
    \begin{minipage}[b]{0.49\textwidth}
        \includegraphics[width=\textwidth]{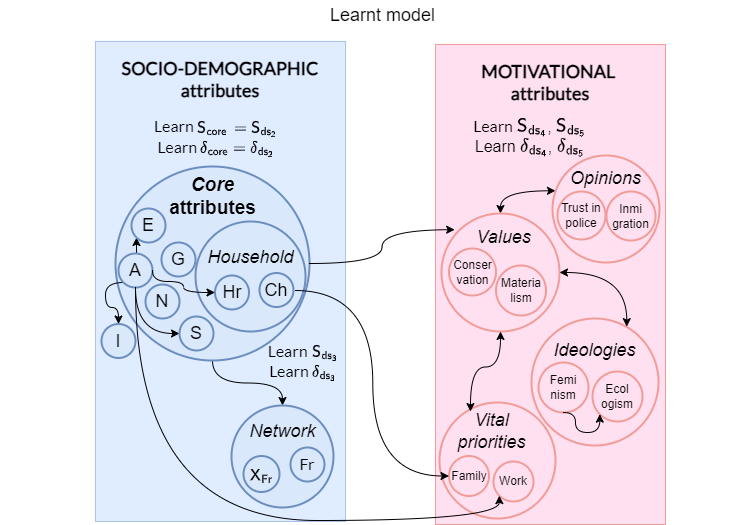}
        \label{fig: S_model}
    \end{minipage}
    \hfill
    \begin{minipage}[b]{0.49\textwidth}
        \includegraphics[width=\textwidth]{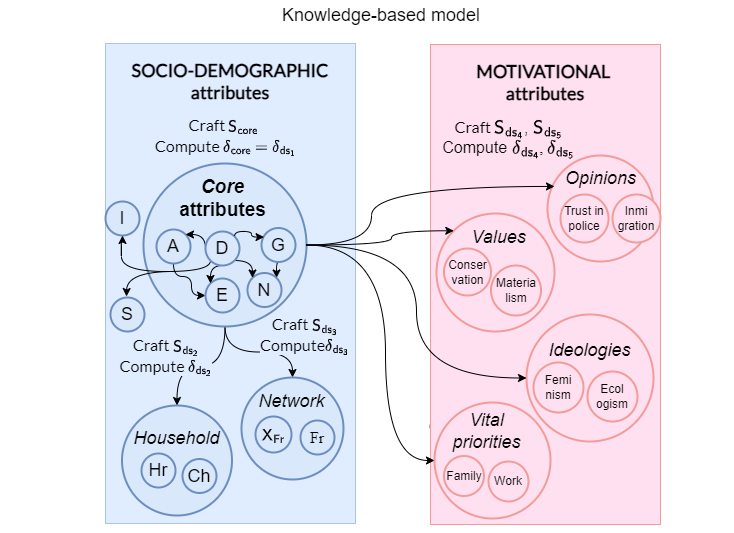}
        \label{fig: H_model}
    \end{minipage}
    \caption{Schematic representation of the learnt and the knowledge-based model structures, learnt or crafted for each specific data source. The diagrams illustrate several interdependencies within the socio-demographic and motivational layers, as well as outer dependencies between them. Double-headed arrows indicate bidirectional dependencies, meaning that the influence can be mutual between connected attribute types. Note that these diagrams do not depict all the identified connections; only a selection of them is shown for simplicity and clarity.  } 
    \label{fig: models}
\end{figure*}

The learnt model draws data from the four microdata sources specified in Table~\ref{tab: data}. It learns the structures and parameters using the hill climb search method and the expectation maximization estimator~\cite{pgmpylearningstructure}. Following the procedure explained in Section~\ref{sec: model selection methodology}, we identify the \textit{core} attributes present in the four datasets. Among the selected datasets, IPUMS stands out as the richest due to its significantly larger interviewed population. Consequently, we learnt both $S_{core}$ and $\delta_{core}$ from it. Once this core is established, we add the remaining attributes from the other datasets and learn their structure and parameters, along with their connection with the core attributes. The final learnt model structure is represented in Fig.~\ref{fig: models}. The structures and parameters $S_{\textit{$ds_{3}$}}$, $\delta_{\textit{$ds_{3}$}}$ and $S_{\textit{$ds_4$}}$ $\delta_{\textit{$ds_4$}}$ encapsulate the dependencies and distributions of the network and motivational attributes, respectively. Furthermore, the remaining structures and parameters $S_{\textit{$ds_{5}$}}$, $\delta_{\textit{$ds_{5}$}}$ can be added as aggregated motivational attributes at the collective level, with a regional scope. 

The dependencies identified by the learnt model align with our expectations regarding the connections between profile attributes, indicating that the model is functioning correctly. For instance, within the socio-demographic layer, age emerges as the predominant influencer, as most of the other attributes depend on it. The social network attributes also exhibit a clear structure where friends share similar traits: the characteristics of individuals are found to influence the characteristics of their friends, as observed with attributes such as gender, education, and age. The individual and collective motivational structures, $S_{\textit{$ds_{4}$}}$ and $S_{\textit{$ds_{5}$}}$, showcase several connections between attributes within the motivational layer and from the socio-demographic one. In the first case, the interdependencies are classified within ideologies, opinions, values and all of them together. such as the parents' ideologies influencing the individual one It is important to highlight that these may vary significantly depending on the city of study, especially those related to politics and trust with institutions. 

For a more concrete description of the whole structure learnt by the model, beyond the schematic one provided in Fig.\ref{fig: models}, we outline some of the interdependencies (connections between attributes within the same layer) and outer dependencies (connections between attributes from the socio-demographic to the motivational layer). Note that, in the context of dependencies, $ a \prec b$ represents the causal relation "$b$ depends on $a$". \\ 

\underline{\textbf{Interdependencies}}
\begin{enumerate}

    \item \textbf{Socio-demographic layer}

\begin{itemize}[noitemsep]
    \item Age  $\prec$  other main demographic attributes
    \item Within network: individual's demographic attributes  \(\prec\) friend's demographic attributes
\end{itemize}

\item \textbf{Motivational layer}
\begin{itemize}[noitemsep]
\item Within ideologies: parents' ideologies $\prec$ individual ideology
\item Within opinions: trust in the monarchy $\prec$ trust in other institutions, opinions about feminism
$\prec$ opinion towards ecologism and multiculturalism 
\item Within Schwartz's values: benevolence $\prec$ universalism, self-direction 
$\prec$ stimulation $\prec$ conformity $\prec$ hedonism $\prec$ power  $\prec$ achievement
\item Within ideologies, values and opinions: Inglehart's index $\prec$ political ideology, religion $\prec$ tradition, opinion towards immigration $\prec$ security and social trust in people
\end{itemize}

\end{enumerate}

\underline{\textbf{Outer dependencies}} 
\begin{itemize}[noitemsep]
    \item Between socio-demographic layer and values: nationality $\prec$  religion and opinion about political independence movements 
, children in household
$\prec$ hedonism
\item Between socio-demographic layer and vital priorities: employment status
$\prec$ satisfaction with professional and economic aspects, age
$\prec$ importance given to work, children in the household
$\prec$ importance given to family and social trust in people 

\end{itemize}

\subsubsection{Knowledge-based Model}
\label{sec: H}

The knowledge-based model draws data from all the sources listed in Table~\ref{tab: data}. We establish a basic structure based on prior knowledge, which involves imposing dependencies, available in data, that seem naturally evident between the attributes. We acknowledge the biases that the knowledge-based structure can introduce, but bear in mind that both models defined are intended to complement each other in the validation step. By detecting which model most accurately represents a specific data source, we plan to select that particular structure and parameters for each data source.

The crafted structure is simpler than the learnt one in terms of connections. The \textit{core} attributes (mostly corresponding to main socio-demographic ones) are chosen to influence all the other socio-demographic and motivational attributes. In fact, the structure is geographically rooted: all attributes are influenced by the territorial unit if there is data that allows us to impose so. The final knowledge-based model structure is represented in Fig.~\ref{fig: models}, where the \textit{core} attributes are the parents of all other attributes. The parameters $\delta_{ds_1}$ are computed using the maximum likelihood method, while the other parameters are estimated using expectation maximization. By manually crafting this structure, rather than learning it (feasible only with microdata sources), we can integrate macrodata from $ds_1$ into the socio-demographic layer. This approach allows us to leverage macrodata that can be more representative than microdata. 

\section{Conclusions and Future Work}
\label{sec: concl}
In this paper, we have presented a novel population synthesis framework that incorporates motivational attributes into agent profiles. By feeding from social survey data, our framework connects information about individuals' values, ideologies, opinions and vital priorities to the rest of their socio-demographic attributes, preserving the representativeness of the population. We propose two different models for evaluation: learnt-based and knowledge-based. These structures lead to the generation of synthetic populations with highly detailed motivational attributes at different geographic scopes. Researchers can use these datasets to initialize their simulations with comprehensive agents' profiles and enhance decision-making architectures. 

 Future work includes developing a hybrid model that leverages the strengths of both the learnt and knowledge-based models: capturing complex variable dependencies effectively while integrating representative and up-to-date macrodata datasets. To achieve this, we will validate the generated synthetic populations against all data sources and compare the performance of the different structures to determine which model's structure most accurately represents each data source selected for the use case. Additionally, the validation should be compared with other emerging machine-learning methods~\cite{VAE1, aemmer2022generative, albiston2024neural}. Once validated, the synthesized population will be integrated into an agent-based model application. By either creating or extending an already existing decision-making architecture with detailed motivational attributes, we are aiming to model the behaviour of agents closer to real-life complex scenarios. 
 
 While agent-based social simulations are always a conceptual simplification of a given social context, their use to inform policy making in sensitive and urgent topics such as poverty mitigation~\cite{Aguilera2024} or public health crisis~\cite{Dignum2020} call for a more nuanced analysis and reflection regarding the values that guide the agents' behaviour. Additional future steps will include testing the replicability of the model in other regions with alternative datasets. The article opens the door to operationalize value alignment of agent-based simulations in different contexts and geographical locations as well as to explore how a diversity of values, ideologies, opinions and vital priorities can affect the effectiveness of policy making by conditioning agents' behaviours. 
\balance
\ack
    
This research has been supported by the EU-funded VALAWAI (\#~101070930), the Spanish-funded VAE (\#~TED2021-131295B-C31) and the Rhymas (\#~PID2020-113594RB-100) projects. Special thanks to Raül Tormos, Head of Methodology and Research at CEO (\textit{Centre d'Estudis d'Opinió}) Generalitat de Catalunya. 

\bibliography{ecaifinal}
\end{document}